\documentclass[reprint,prl,twocolumn,showpacs,superscriptaddress,aps,longbibliography]{revtex4-2}
\usepackage{amsmath}
\usepackage{epstopdf}
\usepackage{amsmath}
\usepackage{amssymb}
\usepackage{amsfonts}
\usepackage[colorlinks=true,citecolor=blue]{hyperref}
\usepackage{graphicx,graphics}

\DeclareGraphicsExtensions{.png,.jpg,.eps}
\usepackage{xcolor}
\usepackage{graphicx}
\usepackage{bm}
\usepackage{hyperref}
\usepackage{braket,bbold,color}
\usepackage{tcolorbox}
\usepackage{ifthen}
\usepackage{dsfont}
\usepackage{tikz}
\usepackage{MnSymbol}
\usepackage{graphicx}
\usepackage{placeins}

\newcommand{\Ham}{\mathcal{H}}

\newcommand{\qsk}[2]{\left\langle #1 \middle| #2 \right\rangle}

\newcommand{\qex}[3]{\left\langle #1 \middle| #2 \middle| #3 \right\rangle}

\begin{document}


\title{Phase diagram of the $J_1$-$J_2$ Heisenberg second-order topological quantum magnet}

\author{Pascal M. Vecsei}
\affiliation{Department of Applied Physics, Aalto University, 00076 Aalto, Finland}

\author{Jose L. Lado}
\affiliation{Department of Applied Physics, Aalto University, 00076 Aalto, Finland}

\date{\today}

\begin{abstract}
Competing interactions in quantum magnets lead to a variety of emergent states, including ordered phases, nematic magnets and
quantum spin liquids. Among them, topological quantum magnets represent a promising platform
to create topological excitations protected by the bulk many-body excitation gap. 
Here we establish the phase diagram of a breathing frustrated antiferromagnetic $J_1$-$J_2$-Heisenberg model,
featuring both ordered states and a higher-order topological quantum magnet state.
Using exact many-body methods based on neural network quantum states and tensor networks, 
we determine the existence of a first order phase transition between stripe order and the topological
quantum magnet and the second order phase transition between the Néel order and quantum magnet phase,
further corroborated by calculations of the many-body gap.
Using an auxiliary fermion parton formalism, we show
the emergence of topological spinon corner modes stemming from the breathing order parameter of the parent Heisenberg model.
Our results establish the breathing frustrated square lattice Heisenberg model as a paradigmatic system to engineer topological quantum magnetism, as recently realized in Ti lattices at MgO.
\end{abstract}

\maketitle

Quantum magnets\cite{Savary2016,RevModPhys.89.025003} represent one of the most exotic states of matter in condensed matter physics.
Frustrated magnetism is a fertile platform to give rise to unconventional excitations,
including magnons\cite{coldea2002direct, schulenburg2002macroscopic}, spinons\cite{Ruan2021}, triplons\cite{Chen2021,PhysRevLett.131.086701}, visons\cite{huh2011visonstates}
and gauge excitations.
Models of quantum magnets have attracted much attention\cite{PhysRevX.14.021053}, 
in particular with regards
to the different unconventional phases that may appear in their phase diagram.
Frustrated Heisenberg models often have a variety of competing phases,
including spin liquid\cite{PhysRevB.92.041105,PhysRevLett.109.067201,Yan2011,PhysRevLett.123.207203,PhysRevX.10.021042}, stripe and zigzag antiferromagnetic\cite{Li2021,PhysRevB.93.214431,Baum2019,PhysRevResearch.2.033011},
valence bond states\cite{PhysRevB.102.014417,PhysRevLett.113.027201,PhysRevLett.121.107202,2024arXiv240617417H}, 
and spin spiral phases\cite{PhysRevLett.82.3899,PhysRevB.89.020408}.
A variety of natural material host potentially frustrated spin models, including
RuCl$_3$\cite{banerjee2017neutronscattering, yokoi2021halfinteger, czajka2021oscillations, yang2022magnetic, jiang2019fieldinduced, janssen2020magnon}, 1T-TaS$_2$\cite{law20171ttas2, maasValero2021quantum, pal2022understanding, li2022fractionalization}, FeSe\cite{she2018quantumspinliquid, gong2017possiblenematic} and Herbertsmithite\cite{norman2016colloquium, freedman2010sitespecific, han2016correlated, wulferding2010interplay, zhang2020variational} are among some of the proposed materials that
may host a variety of competing magnetic states driven by frustration
and quantum fluctuations.

\begin{figure}[t!]
    \includegraphics[width = 1.0\columnwidth]{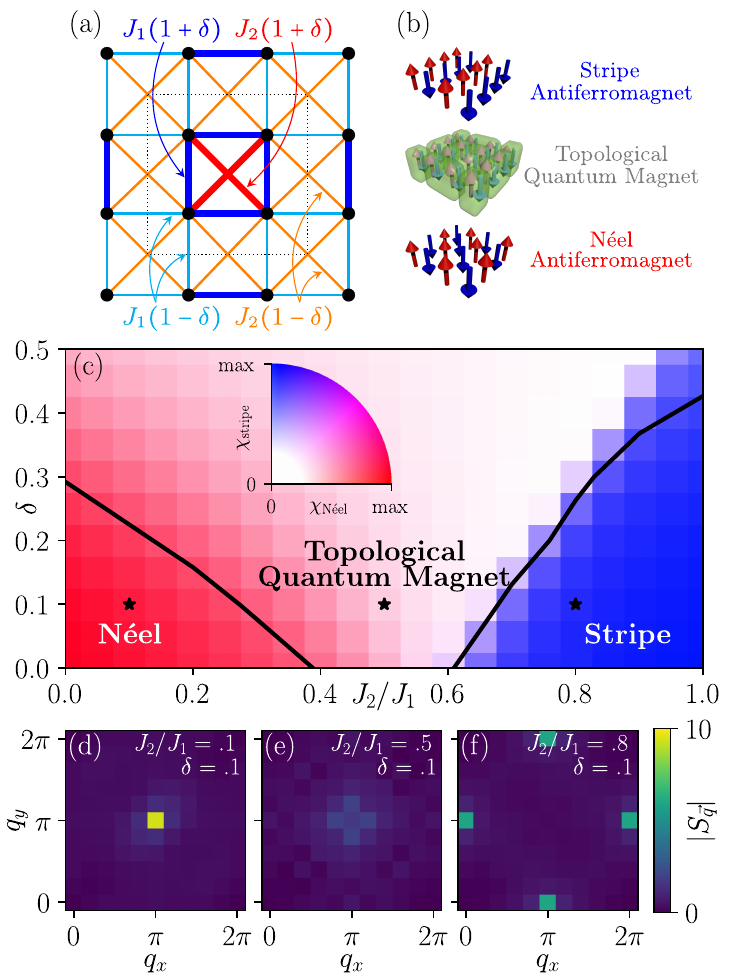}
    \caption{\label{fig:intro}(a,b) Schematic of the breathing $J_1-J_2$ Heisenebrg model, featuring three phases including Neel, stripe and
    Dirac quantum magnet. Panel (c) shows the phase diagram, where the background shows the Néel as red and the stripe order parameters as blue. The lines indicate the phase transition, as obtained using neural network quantum states. Panels (d-f) show the spin structure factor at the marked stars in (c).}
\end{figure}

Beyond naturally occurring materials, artificial platforms based on atomically
precise manipulation have risen as an alternative to create artificial quantum magnets with highly controllable
interactions\cite{RevModPhys.91.041001,Khajetoorians2019,Chen2022}. 
Atomically engineered quantum magnets
with scanning tunneling microscopy
allowed to create models of magnetic criticality\cite{Toskovic2016},
magnets featuring magnon\cite{Spinelli2014},
triplon\cite{PhysRevLett.131.086701,2024arXiv240213590Z}
and spinon excitations\cite{2024arXiv240810045Z,2024arXiv240808801S,2024arXiv240702142S,2024arXiv240808612Y},
and first order and second order topological quantum magnets\cite{Mishra2021,2024arXiv240213590Z,wang2024realizing}.
Such many-body symmetry protected topological phases represent one of the frontiers in
topological matter\cite{peng2021deconfined, you2018higherorder, dubinkin2019higherorder, bibo2020fractional, song2017topological, wang2018symmetric, yoshida2015bosonic, Zeng2019symmetry, kariyado2018Znberry, takayoshi2016fieldtheory, zhang2017unconventional, chen2011twodimensional, wen2017colloquium}.
In particular, the robustness and nature
of the recently realized
second order topological quantum magnet\cite{wang2024realizing}
remains an open problem in quantum magnetism.

In this Letter we establish the phase diagram of the first higher-order topological magnet realized experimentally,
which consists of a frustrated breathing square lattice $J_1-J_2$ Heisenberg model.  We demonstrate the existence of three
widely different regions, including a Néel ordered regime, a stripe antiferromagnetic region, and
a gapped topological quantum magnet as realized experimentally. Our phase diagram is demonstrated using a combination
of exact quantum many-body methods based on tensor-networks and neural-network quantum states, which allows
to trace out the phase diagram both using susceptibilities and quantum many-body gaps.
Using an auxiliary fermion formalism we further show how the exact many-body results are related
to the low energy spinon model of the topological quantum magnet. Our results establish the phase diagram
of a paradigmatic model featuring topological quantum magnetism, establishing the required groundwork
to study exotic phenomena in a realistic system featuring higher-order topological quantum many-body matter.

\begin{figure}
    \includegraphics[width = 1.0\columnwidth]{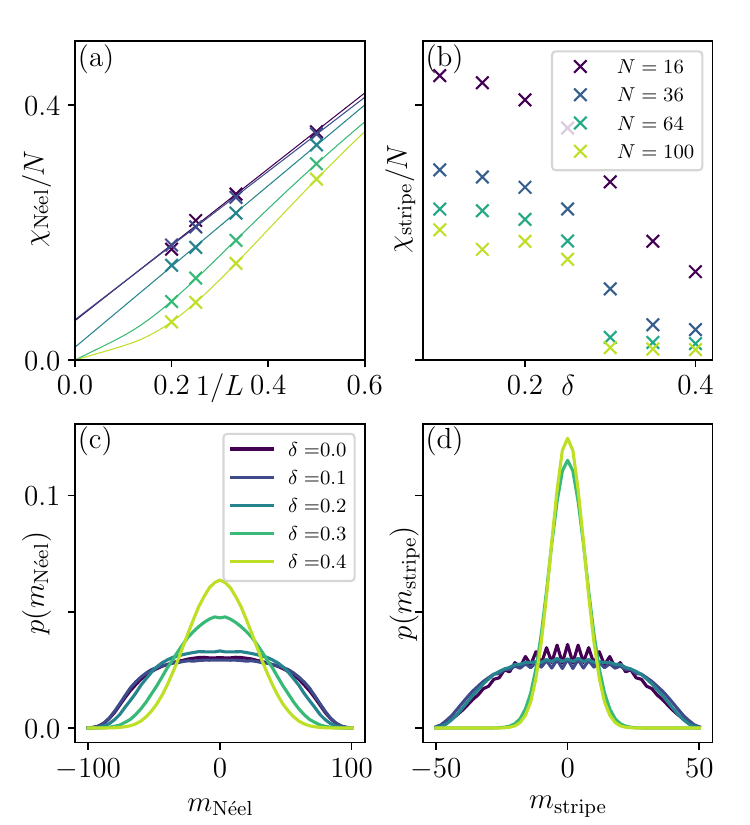}
    \caption{\label{fig:magextrapol}Behavior of the magnetization with increasing system size and breathing $\delta$. Panel (a) shows the shows the second moment of the Néel magnetization as function of $\delta$ and system size $N = 4 L^2$ for fixed $J_2= 0.1$. An extrapolation of form $a/L + b e^{-cL}$ is used to estimate the value in the thermodynamic limit, where $c=0$ if a linear extrapolation yields a positive $b$. Panel (c) shows the distribution of the magnetization for fixed $N=100$. The stripe order phase transition at fixed $J_2 = 0.8$ is treated in panels (b,d). Panel (b) shows the squared stripe order parameter as function of $\delta$, for the different system sizes. A clear jump can be seen between $\delta = 0.25$ and $\delta = 0.3$. This jump is also reflected in a sudden change in the magnetization distribution, as shown in panel (d).}
\end{figure}

\emph{Model:} In the following, we study a breathing antiferromagnetic Heisenberg model on a square lattice with first and second neighbor interactions as realized in Ti at MgO lattices\cite{Willke2018,PhysRevLett.119.227206,Yang2021,wang2024realizing}. The Hamiltonian is
\begin{equation}
    \hat \Ham = \sum_{ij} J_{ij} \vec S_i \cdot \vec S_j,
\end{equation}
where $J_{ij}$ are the interactions and $\vec S_i = (S_i^x, S_i^y, S_i^z)$ are the spin operators used to represent the local spin-1/2 operators as $S_i^k = \sigma_i^k/2$. The relative strength of the interactions is depicted in Fig.~\ref{fig:intro}(a). The model consists of square plaquettes within which the interactions are strong, while between different plaquettes the interaction is weak. The unit cell is indicated by a thin dotted line in Fig.~\ref{fig:intro}(a), and couplings of different strength are marked with different colors. By setting $\delta = 0$, we recover the well-known square lattice $J_1$-$J_2$ antiferromagnetic Heisenberg model. 
At zero breathing $\delta=0$, signatures of a valence bond solid phase (VBS) for $J_2/|J_1| \lessapprox 0.6$, and
a potential gapless Dirac spin-liquid in a region $0.49 \lessapprox J_2/|J_1| \lessapprox 0.54$
have attracted many efforts\cite{richter2010spin, nomura2020diractype, liu2024tensornetwork, ferrari2020gapless, qian2024absence, sindzingre2010phase}. Here we map out the phase diagram
for finite breathing $\delta > 0$ as recently
realized in experiment\cite{wang2024realizing}, further showing how its low energy description
is linked to the previous phases at $\delta=0$.
This quantum magnet realizes a topological phase featuring
a topological degeneracy associated with gapless corner excitations.
We show that the low energy Hamiltonian
can be described with an auxiliary fermion formalism whose effective parameters
can be extracted from our many-body calculations.

\emph{Variational many-body formalism:} We solve two-dimensional many-body 
spin models of up to 100 quantum spins with periodic boundary conditions using Neural
Network Quantum States (NNQS)\cite{carleo2017solving} and matrix product states (MPS)\cite{schollwock2011thedensity}. 
The neural network quantum states we used are group convolutional neural networks (GCNN), which deliver high-accuracy ground states for frustrated models\cite{roth2023highaccuracy, roth2021group}. The GCNN we used are 10 layers deep and have 10 filters per layer. They are fully symmetric under parity and the space group symmetries of the lattice, which has, compared to the square lattice model at $\delta = 0$, a reduced translational symmetry. The nets have up to 361100 complex parameters for $N=100$ spin sites. 
The wave functions were optimized using an
efficient stochastic reconfiguration algorithm\cite{rende2024simple, chen2024empowering,vicentini2022netket}\footnote{Calculations were run as single GPU calculations on half a AMD MI250X each. States were first optimized at specific locations in parameter space, and later annealed to other locations using the pretrained weights to obtain a denser grid.}.
MPS calculations are performed with a maximum bond dimension $\chi_m = 1000$, wavefunctions are optimized with
the density matrix renormalization algorithm\cite{PhysRevLett.69.2863} using quantum number conservation\cite{itensor}.

\emph{Phase Diagram}: We distinguish between three wide phases (cf.~Fig.~\ref{fig:intro}(b)): a Néel antiferromagnetic for small $J_2$ and $\delta$, an intermediate topological quantum magnet phase, and a stripe order antiferromagnetic phase for large $J_2$ and small $\delta$. The order parameters to distinguish
between these three phases are defined through the operators 
\begin{equation}
    \hat M_\text{Néel/stripe} = \sum_{\vec \alpha} \zeta_{\vec \alpha} \hat S_{\vec \alpha}^z,
\end{equation}
where the sum runs over the two dimensional indices $\vec \alpha = (i,j)$, and $\zeta_{\vec \alpha} = 2(-1)^i (-1)^j$ for the Néel order parameter and $\zeta_{\vec \alpha}  = ((-1)^{i} + (-1)^{j})$ for the stripe order parameter. As the model is time reversal symmetric, the first moment of these order parameters vanishes.
The second moment reflects the Neel and stripe susceptibilities, and allows us to distinguish ordered and disordered phases. 
The resulting phase diagram is shown in Fig.~\ref{fig:intro}(c) as the solid black lines, 
where the background showing the Neel and stripe magnetic susceptibilities for a 36 spin system
obtained from matrix product state calculations with periodic boundary conditions. 
We show in  Fig~\ref{fig:intro}(d)-(f) the spin structure factor $S(\vec q) = 4 \sum_{\vec \alpha \vec \beta} e^{-i \vec \beta \cdot \vec q / 2L} \langle \vec S_{\vec \alpha} \cdot \vec S_{\vec \alpha + \vec \beta} \rangle / N $ for the three points marked with stars in Fig.~\ref{fig:intro}(c). Panel Fig~\ref{fig:intro}(d) is located well within the Néel phase, featuring
a clear peak at $(q_x, q_y) = (\pi,\pi)$ signaling the Neel order. Fig~\ref{fig:intro}(e) 
is located within the quantum magnet phase, and the magnetic structure factor shows no clear peak. The stripe order antiferromagnet is shown in panel Fig~\ref{fig:intro}(f), featuring peaks at the locations $(q_x, q_y) = (\pi,0)$ and $(0,\pi)$ as expected from a stripe phase.

To locate the phase transition, we study the magnetization as a function of system size, breathing $\delta$ and next-to-nearest neighbor exchange $J_2$. In Fig.~\ref{fig:magextrapol}(a), we show the second moment of the magnetization, which increases as we move into the symmetry broken Néel antiferromagnetic phase, for fixed $J_2 = 0.1$. We evaluate this quantity from our numerical calculations for differently sized finite clusters, and then extrapolate linearly to the thermodynamic limit. A non-zero extrapolated value of $\chi_\text{Néel}/N = \llangle M_\text{Néel}^2 \rrangle / N^2$ implies that the susceptibility $\chi_{\text{Neel}} = \llangle M_\text{Néel}^2 \rrangle / N$ diverges, as expected at the phase transition. In Fig.~\ref{fig:magextrapol}(c) we show the distribution of the Neel magnetization for $J_2 = 0.1$ and variable $\delta$ for fixed system size $N=100$ quantum spins. It is defined as $p(m_\text{Néel/stripe}) = \qex{\psi}{P_{m_\text{Néel/stripe}}}{\psi} / \qsk{\psi}{\psi}$, where $P_{m_\text{Néel/stripe}}$ is the projector into the eigenspace of $\hat M_\text{Néel/stripe}$ with eigenvalue $m_\text{Néel/stripe}$ . While for large $\delta$, in the quantum magnet phase, this distribution is Gauss-like in shape and clearly peaked at $m_{z, \text{Néel}}= 0$, as the breathing decreases the distributions becomes wider. This phase transition from the quantum magnet to the Néel antiferromagnetic phase shows the typical behavior of a second order phase transition, with a continuous change in the order parameter for finite system sizes. 

Next, let us study the transition into the stripe-antiferromagnetic phase. There, we encounter a first order phase transition, which is signaled by a rapid change in the magnetization. This can be also observed as a kink in the ground state energy, a change in the degeneracy of the ground state already on finite clusters, and a change in the symmetry of the lowest $S_z = 1$ state. The sudden change in the magnetization is reflected in the second moment, which is shown in Fig.~\ref{fig:magextrapol}(b) as a function of $\delta$ and system size. For large system sizes, there is a clear jump in this quantity at the phase transition, which allows us to locate the phase transition. In Fig.~\ref{fig:magextrapol}(d), we show the stripe magnetization distribution for $J_2/J_1 = 0.8$ for varying $\delta$. While for large $\delta$, in the quantum magnet phase, the distribution is strongly peaked at around zero, it suddenly becomes much wider as we pass through the pass transition.

\begin{figure}
    \includegraphics[width = 1.0\columnwidth]{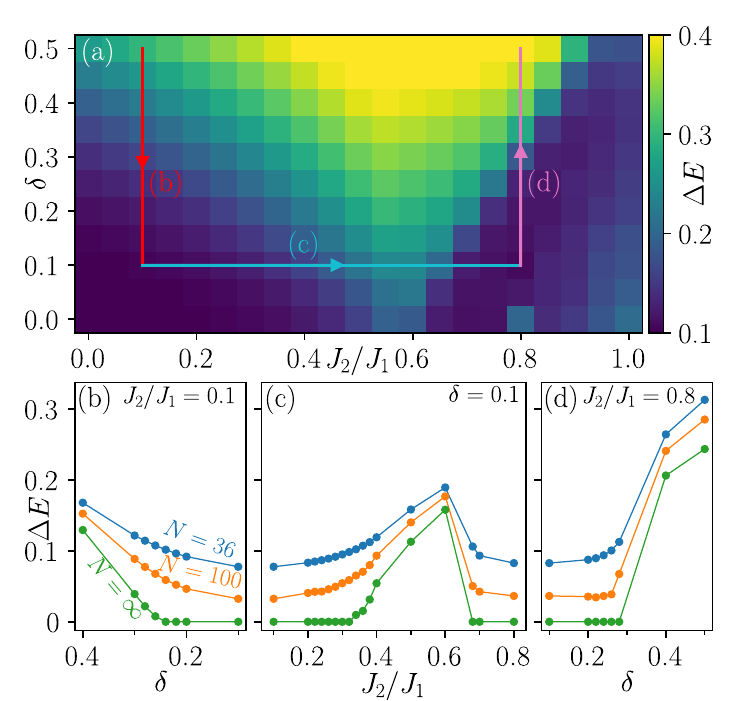}
    \caption{\label{fig:energy}The background color in panel (a) shows the gap between the lowest $S_z = 1$ state and the ground state, from $N=36$ system by MPS. We clearly see two regions with low gap, which are the Néel and the stripe phase regions, both of which have gapless magnons. The gap along the path depicted in panel (a) is shown in panels (b)-(d), with data obtained from NNQS and a linear extrapolation of these two data points.}
\end{figure}

\emph{Phase transition from many-body gaps}: 
In the antiferromagnetic phases, both the Néel and the stripe order, gapless magnons exist above the ground state. In contrast, in the topological quantum magnet phase for $\delta>0$ with closed boundary conditions, the system features a finite gap in the thermodynamic limit. 
The transition between those states can be 
detected by looking for a closing of the gap between the ground state and the $S_z=1$ excitations. To do this, we map out the gap between the $S_z=0$ and $S_z =1$ sectors for a $N = 36 $ spin system, as shown in Fig.~\ref{fig:energy}(a). We see two regions with a clearly smaller gap, corresponding to the Néel and stripe antiferromagnet. To obtain more detailed information, we study the gap with NNQS along the path shown in Fig.~\ref{fig:energy}(a). Fig.~\ref{fig:energy}(b) shows the vertical section for fixed $J_2 = 0.1$, passing from the topological quantum magnet to the Néel ordered phase. We show the gap for system sizes $N=36$ and $N=100$, as well as a power law extrapolation to the thermodynamic limit\footnote{Gaps are fit to a functional form $\Delta(N) = a + b/\sqrt{N}$, as expected for linearly dispersive magnons.}. As $\delta$ decreases, so does the size of the gap, which eventually reaches zero at a value of $\delta$ in good agreement with the results from the magnetization calculations. Fig.~\ref{fig:energy}(c) shows the gap along the horizontal path, at fixed $\delta = 0.1$. For $J_2 < 0.5$, the gap first stays zero and then increases as the system passes from the Néel ordered into the topological quantum magnet phase. Eventually, as the system transitions into the stripe ordered phase, it becomes gapless again. Fig.~\ref{fig:energy}(d) shows the evolution of the gap along the fixed line at $J_2 = 0.8$. The energies of the excited states were obtained by looking for the lowest energy state at finite $S_z$.\footnote{The evolution of excited states in the $S_z = 1 $ and $S_z = 0$ sectors is shown in the supplemental material, showing a clear peak of the ground state energy at the first order phase transition between stripe order and quantum magnet phases, and the change in the degeneracy of the ground state at the same point.}

\begin{figure}
    \includegraphics[width = 1.0\columnwidth]{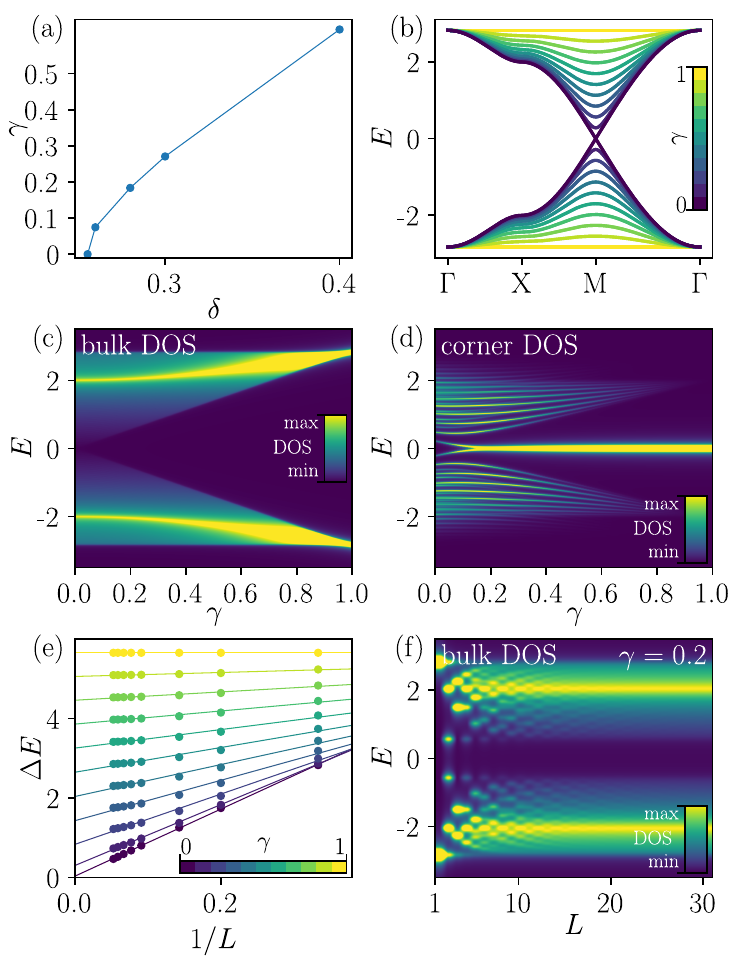}
    \caption{\label{fig:spinonmodel}Plots about the corresponding spinon model. Panel (a) shows the mapping between $\gamma$ and $\delta$ for fixed $J_2 = 0.1$, as obtained by the assumption that the equivalent values have the same ratio $\Delta E_{L=5}/\Delta E_{L=3}$. Panel (b) shows the band structure as function of the breathing $\gamma$. Panel (c) shows the DOS for infinite systems, as function of dimerization $\gamma$. The corner DOS for finite systems of size $20\times 20$ is shown in panel (d) as function of $\gamma$. Panel (e) shows the size-dependent gap size for different values of $\gamma$, with a linear extrapolation. Panel (b) shows the the bulk DOS for finite sized periodic systems as function of the size. }
\end{figure}

\emph{Auxialiary pseudofermion representation:} 
In the quantum disordered regime, the Heisenberg Hamiltonian
$
H = \sum_{ij} J_{ij} \vec S_i \cdot \vec S_j
$
can be approximately solved using with auxiliary
Abrikosov fermions\cite{PhysRevB.44.2664,PhysRevLett.66.1773,Baskaran1987,PhysRevB.65.165113} $\vec S_n = \frac{1}{2} \sum_{s,s'} \vec \sigma_{s,s'} f^\dagger_{n,s} f_{n,s'}$ with the constraint
$ \sum_s f^\dagger_{n,s} f_{n,s} = \mathcal{I}$. 
Using the replacement in the Heisenberg Hamiltonian and performing a
saddle point approximation in the
effective action gives rise to the effective spinon Hamiltonian

\begin{equation}
\mathcal{H} = \sum_{ij}\Gamma_{ij} f^\dag_{i,s} f_{j,s},
\end{equation}

where the spinon hoppings give rise to a $\pi$-flux model $\prod_{ij} \text{sign}(\Gamma_{ij}) = -1$ around each plaquette, and $\Gamma_{ij} = \Gamma_0 (1 \pm \gamma)$ featuring a
first neighbor breathing analogous to Fig.~\ref{fig:intro}a. 
The previous Hamiltonian with $\gamma=0$ describes the gapless Dirac spin liquid\cite{nomura2020diractype, ferrari2020gapless} of the $J_1$-$J_2$ Heisenberg model, featuring gapless Dirac spinons.
The breathing order parameter $\gamma$ opens a gap in the spinon spectra,
leading to a state analogous to the plaquette bond ordered phase\cite{PhysRevB.74.144422,PhysRevB.102.014417,PhysRevLett.121.107202,nomura2020diractype,2024arXiv240617417H}.
Within this model, the gapless Dirac spin liquid state of the square lattice corresponds to the 
limit $\gamma=0$, whereas the plaquette ordering stems from a dynamically generated Dirac mass
leading to $\gamma \ne 0$.
The topological nature of the spin many-body breathing phase can be understood in terms of the topological classification arising from the combination of time reversal symmetry $\mathcal T$ with one of the point group symmetries, diagonal reflection $R_{xy}$ or $C_4$ rotation\cite{peng2021deconfined, you2018higherorder, dubinkin2019higherorder}. The combination of these symmetries imposes a topological classification onto the space of many-body gapped bosonic Hamitonians with spin $1/2$. The nontrivial phase, which arises for $\delta < 0$, has topological corner modes with spin 1/2, and is referred to as higher-order symmetry protected topological phase\cite{peng2021deconfined}. For $\delta>0$ we encounter the corresponding trivial phase for which all of the sites are translated diagonally by one site. 
In this gapped phase, the system becomes topologically nontrivial as can be evaluated using the nested Wilson loop of the effective spinon Hamiltonian\cite{Benalcazar2017,PhysRevB.96.245115,Schindler2018,PhysRevResearch.2.022049}, and leading to topological spinon corner modes. 

The parameters of the effective spinon model can be extracted by comparing the many-body gaps of the exact
quantum many-body calculations with those of the effective model.
In particular, we compare the ratio of the gap sizes between systems with $L=5$ and $L=3$, $\Delta E_{L=5}/\Delta E_{L=3}$ between the spinon model and the numerical results for the Heisenberg model at $\delta = 0.1$. This allows us to obtain a mapping between the breathing parameter of the spinon model $\gamma$, and the corresponding breathing parameter of the Heisenberg model, $\delta$, as shown in Fig.~\ref{fig:spinonmodel}(a).
When the breathing becomes non-zero in the effective model, Dirac spinons develop a mass leading to gapped
excitations, as depicted in Fig.~\ref{fig:spinonmodel}(b). Fig.~\ref{fig:spinonmodel}~(c) shows the spinon DOS for the infinite system in the bulk. The corner spinon DOS is shown in Fig.~\ref{fig:spinonmodel}(d), where a zero mode can be clearly seen for large enough breathing.
The gap in the thermodynamic limit can be obtained from an extrapolation from small systems in Fig.~\ref{fig:energy}(e),
in analogy to our exact quantum many-body calculations. It is worth noting that the extrapolation
follows the linear dependence expected from gapped Dirac spinons for small dimerization, 
as shown in Fig.~\ref{fig:spinonmodel}(e).
This phenomenology can be directly observed by computing the bulk spinon DOS as a function of the system size
at fixed $\gamma = 0.2$ as shown in Fig.~\ref{fig:spinonmodel}~(f), showing the decrease and saturation of the
gap.

\emph{Conclusion}
Frustration in quantum many-body systems is a recognized driving force for unconventional phenomena. Here, we have presented the phase diagram of a $J_1$-$J_2$ Heisenberg model on a breathing square lattice,
a recently experimentally realized model in Ti atoms on MgO featuring topological corner modes. Using exact many-body methods based on large-scale tensor network and neural network quantum state calculations, we have mapped out its magnetic phase diagram, revealing the phase boundaries between the Néel antiferromagnetic, the topological quantum magnet and the stripe-antiferromagnetic phases. This phase diagram contains a region with a topological quantum magnet phase,
sharing a first order phase transition to the stripe order and
a second order phase transition to the Neel state. Furthermore, using an auxiliary fermion formalism we have
mapped the results of our many-body methods to a gapped quantum spinon model,
revealing the evolution of the topological bulk gap and topological corner modes in this model.
Our findings put forward a rich phase diagram of a model realizing topological quantum magnetism
as recently demonstrated experimentally with Ti atoms at MgO.
Our results establish a paradigmatic system
to explore topological phases in frustrated breathing Heisenberg models,
and their connection to gapped and gapless spinon phases through neural network and tensor network
many-body methods.

\textbf{Acknowledgements}
We thank C. Flindt, M. Niedermeier, K. Yang, S. Dominguez and T. Antao for useful discussions. We acknowledge the computational resources provided by the Aalto Science-IT project and CSC (Finland) for awarding this project access to the LUMI supercomputer, owned by the EuroHPC Joint Undertaking, hosted by CSC (Finland) and the LUMI consortium. We acknowledge the support from the Research Council of Finland through grants (Grants No.~331342 and No.~358088), the Finnish Quantum Flagship, and the Finnish Centre of Excellence in Quantum Technology (Project No.~312299), and the Jane and Aatos Erkko Foundation.

\bibliography{Ref-Lib}

\end{document}